\documentclass[superscriptaddress,column,showpacs,a4paper,
amssymb,amsmath,nobibnotes,aps,prd,longbibliography,
showkeys,
nofootinbib,notitlepage]{revtex4-1}
\usepackage{bigints}
\usepackage{graphicx}
\usepackage{float}
\usepackage{epsf}
\usepackage{bm}
\usepackage{amsmath}
\usepackage{amsfonts}
\usepackage{amssymb}
\usepackage{epstopdf}
\usepackage{natbib}
\usepackage{color}%
\usepackage{tikz}
\usepackage{rotating}
\usetikzlibrary{decorations.markings,arrows.meta}
\providecommand{\U}[1]{\protect\rule{.1in}{.1in}}
\newcommand{\be}{\begin{equation}}
\newcommand{\ee}{\end{equation}}

\newcommand{\mincir}{\raise
-3.truept\hbox{\rlap{\hbox{$\sim$}}\raise4.truept\hbox{$<$}\ }}
\newcommand{\magcir}{\raise
-3.truept\hbox{\rlap{\hbox{$\sim$}}\raise4.truept\hbox{$>$}\ }}

\usepackage{xcolor}
\usepackage{hyperref}
\hypersetup{colorlinks=true, linkcolor=blue,citecolor=magenta}

\begin{document}

\title{{Novikov Coordinates and the Physical Description of Gravitational Collapse}}

\author{Jaume de Haro}
\email{jaime.haro@upc.edu}
\affiliation{Departament de Matem\`atiques, Universitat Polit\`ecnica de Catalunya, Diagonal 647, 08028 Barcelona, Spain}

\begin{abstract}

We show that the Novikov coordinates can be obtained in a direct and physically transparent way from the radial geodesics of massive particles with negative energy in the Schwarzschild spacetime. These geodesics form a complete congruence that covers the entire spacetime. By rectifying this family of trajectories using the proper time as the time coordinate, the Novikov variables naturally emerge, providing a clear dynamical interpretation of the different regions usually identified as black-hole and white-hole sectors.

In Novikov coordinates, observers at fixed spatial position follow free-fall trajectories. From their perspective, the gravitational collapse of a dust star is completed in a finite proper time, independently of their initial distance from the star. In contrast, observers described by Schwarzschild–Droste coordinates perceive the boundary of the collapsing star as taking an infinite coordinate time to reach the horizon.

We emphasize that Schwarzschild–Droste observers are static with respect to the center of mass of the star and therefore cannot be in free fall. The use of these coordinates implicitly requires the presence of a force that compensates the gravitational attraction. From this viewpoint, the apparent infinite-time collapse is not a physical effect but a coordinate artifact associated with non-inertial observers.

\end{abstract}

\vspace{0.5cm}


\maketitle


\section{Introduction}

The Schwarzschild solution \cite{Schwarzschild} plays a central role in our understanding of black holes. When written in Schwarzschild–Droste (SD) coordinates \cite{Droste}, the spacetime appears to possess a coordinate singularity at the Schwarzschild radius, which obscures the physical interpretation of the horizon and of dynamical processes such as gravitational collapse. 
{

{Immediate attempts to analyse this coordinate singularity were made by Painlevé \cite{Painleve} and Gullstrand \cite{Gullstrand}, who introduced coordinate systems --now called Painlevé–Gullstrand (PG) coordinates-- that are regular at the horizon and adapted to freely falling observers. These coordinates allow one to describe the motion of infalling particles across the horizon in a physically transparent way, but they do not provide a maximal analytic extension of the Schwarzschild spacetime, as they cover only part of the maximally extended manifold (see \cite{Kanai}
 to describe the collapsing star
 in generalized PG coordinates
 not only outside but also inside
the event horizon in a single coordinate patch,  and \cite{Lin, Faraoni}
the problems involved with PG coordinates 
to describe  Reissner-Nordström and Schwarzschild-anti-deSitter spacetimes
).}
{Other horizon-penetrating coordinate systems, such as Eddington–Finkelstein coordinates \cite{Eddington1924, Finkelstein1958}, also provide a regular description at the Schwarzschild horizon. While these systems are important in various applications, they are not discussed here in detail because our focus is on coordinates directly adapted to freely falling observers, which offer the clearest operational interpretation of proper time and radial collapse.}

}
This limitation motivated the introduction of alternative coordinate systems—most notably the Kruskal–Szekeres (KS) and Novikov coordinates (see for instance \cite{Kruskal, Szekeres,  Novikov, Weinberg, MTW, Landau, Lemos})—that allow for a maximal analytic extension of the Schwarzschild spacetime.

The Kruskal–Szekeres coordinates provide a global description in which the line element is manifestly regular at the horizon and conformally flat in the radial–temporal sector. However, their interpretation is often subtle: the Kruskal time does not correspond to the proper time—the invariant introduced by Minkowski in \cite{Minkowski} which corresponds to the time measured by a comoving observer—of any distinguished family of observers, and the extended spacetime includes regions that are commonly interpreted as white holes and parallel asymptotically flat universes. While mathematically consistent, the physical meaning of these regions remains controversial, particularly when one is interested in realistic gravitational collapse rather than eternal black hole solutions.


An alternative and physically transparent approach is provided by the Novikov coordinates. These coordinates are constructed from the congruence of freely falling massive particles and are naturally adapted to the proper time along geodesic motion. As a result, the Novikov time coordinate has a direct operational meaning: it coincides with the time measured by freely falling observers, which is the invariant notion of time in General Relativity. In this framework, the Schwarzschild spacetime is foliated by hypersurfaces orthogonal to free--fall worldlines (geodesics in a more geometric language), and the horizon appears as a regular hypersurface crossed in finite proper time.

 {In the present essay, intended as a pedagogical exposition for graduate students and researchers seeking a physically transparent understanding of gravitational collapse, we build on these considerations to emphasize the operational meaning of proper time and the role of coordinate choices. 
To illustrate these ideas concretely, we derive the Novikov coordinates explicitly from the radial timelike geodesics of the Schwarzschild spacetime, showing how this construction provides a clear dynamical interpretation of the black hole geometry in the idealized case of radial free fall.} 
In particular, we emphasize that the apparent presence of white-hole–like trajectories in the eternal Schwarzschild solution is a consequence of the maximal analytic extension and of time-reversal symmetry, rather than evidence for a physically realizable process. When the physically motivated time orientation is fixed—requiring that the Schwarzschild time increases along freely falling trajectories outside the horizon—the interpretation becomes unambiguous.


We then present a detailed comparison between the Schwarzschild–Droste, Kruskal–Szekeres, and Novikov coordinate systems. Special attention is paid to the role of observers: static observers in Schwarzschild coordinates necessarily experience a non-gravitational force that counterbalances gravity, whereas observers at fixed spatial coordinates in the Novikov frame are in free fall.

Our analysis clarifies the physical content of the eternal Schwarzschild spacetime and highlights the advantages of coordinate systems adapted to free--fall motion.
The results support the view that the essential physics of {the eternal Schwarzschild black hole} is most naturally understood from the perspective of freely falling observers, in accordance with the Equivalence Principle.

Finally, we deal with gravitational collapse {in the specific case of a spherically symmetric dust star}. A crucial aspect in its description is the choice of coordinates and, more importantly, their physical interpretation. Schwarzschild coordinates are adapted to observers that remain at fixed radial distance from the center of the star. However, such observers are not freely falling: in order to stay static with respect to the collapsing matter, even when located arbitrarily far from the star, they must be subjected to an external force that exactly compensates the gravitational attraction. Therefore, Schwarzschild coordinates implicitly describe a non-inertial family of observers, and their use introduces into the problem an additional, non-gravitational interaction that is not part of the physical collapse process itself.

In contrast, Novikov coordinates are constructed from the congruence of freely falling geodesics followed by the dust particles composing the star. The temporal coordinate coincides with the proper time of these particles, providing a direct and physically meaningful description of the collapse across the horizon and up to the formation of the singularity. Within this framework, the collapse is described as a single, irreversible dynamical process, without the need to introduce artificial time delays or external forces. This makes free--falling coordinates particularly well suited for the analysis of gravitational collapse {in the idealized dust and spherical symmetry approximation (see \cite{Abellan2025} for studies of collapse with matter under pressure, and \cite{Balakrishna2015} for collapse including quantum effects)}, allowing one to clearly distinguish genuine physical effects from coordinate-dependent artifacts associated with static, non-inertial observers.

\section{The dynamical equations}

The invariant line element in Schwarzschild--Droste coordinates is \cite{Schwarzschild, Droste}
\begin{eqnarray}
ds^2=\left(1-\frac{2MG}{r}\right)dt^2-
\left(1-\frac{2MG}{r}\right)^{-1}dr^2-r^2d\Omega^2 .
\end{eqnarray}

We are interested in radial trajectories, that is, $d\Omega^2=0$.
Therefore, we consider
\begin{eqnarray}
ds^2=\left(1-\frac{2MG}{r}\right)dt^2-
\left(1-\frac{2MG}{r}\right)^{-1}dr^2 .
\end{eqnarray}

The equations of motion follow from the extremization of the proper time, a principle already highlighted by Planck in Special Relativity \cite{Planck} and subsequently extended by Einstein to include static gravitational fields \cite{Einstein1912b},
\begin{eqnarray}
s=\int ds=\int
\sqrt{
\left(1-\frac{2MG}{r}\right)-
\left(1-\frac{2MG}{r}\right)^{-1}(r')^2
}\,dt
\equiv \int \sqrt{F(r,r')}\,dt ,
\end{eqnarray}
where we have introduced the notation
$r'\equiv\frac{dr}{dt}$.

The Euler--Lagrange equation yields
\begin{eqnarray}
\frac{1}{\sqrt{F}}
\frac{d}{dt}
\left(
\frac{r'\left(1-\frac{2MG}{r}\right)^{-1}}{\sqrt{F}}
\right)
=
-\frac{1}{2}F^{-1}\partial_r F .
\end{eqnarray}

Introducing the proper-time derivatives
$\dot{t}=\frac{dt}{ds}$ and $\dot{r}=\frac{dr}{ds}$,
this equation can be rewritten as
\begin{eqnarray}
\frac{d}{ds}
\left(
\dot{r}\left(1-\frac{2MG}{r}\right)^{-1}
\right)
=
-\frac{MG}{r^2}
\left[
\dot{t}^2+
\left(1-\frac{2MG}{r}\right)^{-2}\dot{r}^2
\right],
\end{eqnarray}
where we have used the identity
$\dot{t}^2=F^{-1}(r,r')$.

The normalization of the four-velocity implies the constraint
\begin{eqnarray}\label{constraint}
\left(1-\frac{2MG}{r}\right)\dot{t}^2-
\left(1-\frac{2MG}{r}\right)^{-1}\dot{r}^2=1 .
\end{eqnarray}

Using this constraint, the equation of motion reduces to the
Newtonian form
\begin{eqnarray}
\ddot{r}=-\frac{MG}{r^2},
\end{eqnarray}
whose first integral is
\begin{eqnarray}
\dot{r}^2=\frac{2MG}{r}+2E ,
\end{eqnarray}
where  the {\it energy} $E$ is an integration constant. {Recall that negative values of $E$ simply correspond to gravitationally bound
radial trajectories, exactly as in the Newtonian theory. In the present
framework, the dynamics is formulated entirely in terms of the proper time
of the freely falling particle, and the equation of motion coincides exactly
with Newton's equation for a unit mass moving in the gravitational field
$-MG/r$.

Accordingly, $E$ represents the mechanical energy per unit mass of the
radial motion, defined with respect to the particle's proper time, and its
sign has the same physical interpretation as in classical mechanics. Negative
values of $E$ therefore do not signal any physical pathology, but simply
describe bound trajectories. 


}

Substituting this expression into Eq.~(\ref{constraint}), we obtain
\begin{eqnarray}
\left(1-\frac{2MG}{r}\right)^2\dot{t}^2=1+2E .
\end{eqnarray}

Therefore, the radial geodesics parametrized by the proper time satisfy
\begin{eqnarray}
\dot{r}^2=\frac{2MG}{r}+2E,
\qquad
\left(1-\frac{2MG}{r}\right)^2\dot{t}^2=1+2E .
\end{eqnarray}

The following remarks are in order:

\begin{enumerate}
\item
The same equations can be obtained by extremizing the action
\begin{eqnarray}
s=\int ds
=\int
\left[
\left(1-\frac{2MG}{r}\right)\dot{t}^2-
\left(1-\frac{2MG}{r}\right)^{-1}\dot{r}^2
\right] ds
\equiv \int L\,ds ,
\end{eqnarray}
together with the constraint (\ref{constraint}), namely $L=1$.

{\item 
The variational principle employed here is based on the extremization of
the proper time and therefore leads to a Lagrangian depending only on the
coordinates and their first derivatives. This is not an additional
assumption, but a direct consequence of the geometric nature of the
invariant line element $ds^2$, which fully characterizes the motion of a
structureless massive particle.

Lagrangians involving higher--order derivatives would describe systems
with additional internal degrees of freedom or nonlocal interactions and
are not appropriate for the description of free fall in a gravitational
field. In particular, such higher--order terms would not correspond to
geodesic motion and would obscure the physical interpretation of proper
time as the fundamental dynamical parameter.

}

\item

The dynamical equation $\ddot{r}=-\frac{MG}{r^2}$ coincides with the
Newtonian equation of motion when the coordinate time is replaced by
the proper time. Therefore, an alternative way to recover the invariant
line element is to solve the inverse problem: to determine a Lagrangian
of the form
\begin{eqnarray}
L=A(r)\dot{t}^2-B(r)\dot{r}^2 ,
\end{eqnarray}
such that the Euler--Lagrange equations reproduce the Newtonian dynamics
with respect to the proper time. Imposing the constraint $L=1$, one
finds
\begin{eqnarray}
A(r)=B^{-1}(r)=1-\frac{2MG}{r}.
\end{eqnarray}

\item

For a massive particle of mass $m$ moving radially under the gravitational field generated by a central mass $M$, the equation of motion can be written as
\begin{eqnarray}
    -m\ddot{r}-\frac{mMG}{r^2}=0.
\end{eqnarray}
This equation is nothing but the relativistic version of  Newton's second law for a particle subject to the gravitational attraction of the mass $M$.

Introducing the inertial force
\begin{eqnarray}
    F_{\rm i}\equiv -m\ddot{r},
\end{eqnarray}
and identifying the gravitational force acting on the particle as
\begin{eqnarray}
    F_{\rm g}\equiv -\frac{mMG}{r^2},
\end{eqnarray}
the equation of motion can be rewritten, in the spirit of D'Alembert's principle \cite{Dalembert}, as
\begin{eqnarray}
    F_{\rm i}+F_{\rm g}=0.
\end{eqnarray}

From this point of view, no net force acts on the particle: the inertial force exactly compensates the gravitational one. This dynamical balance provides a clear physical interpretation of free fall and is fully consistent with Einstein's original formulation of the Equivalence Principle, according to which an observer in free fall does not feel its own weight.
In this sense, gravity does not disappear in free fall; rather, it is dynamically balanced by the inertial response of the particle itself.

\end{enumerate}


\subsection{Solution of the dynamical equations}

We start by considering the solutions with null energy, i.e.,  $E=0$. 
We select trajectories that fall toward $r=0$, satisfying
\begin{eqnarray}
\dot{r}=-\sqrt{\frac{2MG}{r}},
\qquad
\left(1-\frac{2MG}{r}\right)\dot{t}=1,
\end{eqnarray}
so that, outside the Schwarzschild horizon, the Schwarzschild time coordinate $t$ of a static observer evolves in the same direction as the proper time of the freely falling particle.
This choice ensures that, outside the horizon, both the proper time of the particle and the Schwarzschild time increase monotonically along the trajectory.

The solution with initial conditions
\begin{eqnarray}
0=t_0+2\sqrt{2MG\,r_0}
+2MG\ln\left|
\frac{\sqrt{r_0}-\sqrt{2MG}}{\sqrt{r_0}+\sqrt{2MG}}
\right|,
\qquad r_0>2MG,
\end{eqnarray}
is given by
\begin{eqnarray}\label{worldlines}
s&=&t(s)+2\sqrt{2MG\,r(s)}
+2MG\ln\left|
\frac{\sqrt{r(s)}-\sqrt{2MG}}{\sqrt{r(s)}+\sqrt{2MG}}
\right| ,\nonumber\\
\frac{2r_0^{3/2}}{3\sqrt{2MG}}
&=&t(s)+\frac{2}{\sqrt{2MG}}
\left(
\frac{r^{3/2}(s)}{3}+2MG\sqrt{r(s)}
\right)
+2MG\ln\left|
\frac{\sqrt{r(s)}-\sqrt{2MG}}{\sqrt{r(s)}+\sqrt{2MG}}
\right|.
\end{eqnarray}

Note that $r(s)$ has a simple expression in terms of the proper time $s$
and the initial position $r_0$,
\begin{eqnarray}
r(s)=\left(
r_0^{3/2}-\frac{3}{2}\sqrt{2MG}\,s
\right)^{2/3}.
\end{eqnarray}

The particle reaches the horizon $r=2MG$ at the finite proper time
\begin{eqnarray}
s_{\rm H}=\frac{2}{3\sqrt{2MG}}
\left(
r_0^{3/2}-(2MG)^{3/2}
\right),
\end{eqnarray}
while the corresponding coordinate time diverges,
$t(s_{\rm H})=+\infty$.
Nevertheless, the worldline can be extended for $s>s_{\rm H}$ up to
\begin{eqnarray}
s_{{\rm H}}=\frac{2}{3\sqrt{2MG}}\,r_0^{3/2},
\end{eqnarray}
when the particle reaches the point $(t=s_{{\rm H}},\, r=0)$.

Therefore, we obtain the following bijection:
\begin{eqnarray}
\left(
s,\;\frac{2r_0^{3/2}}{3\sqrt{2MG}}
\right)
\longleftrightarrow
\left(t(s),\,r(s)\right).
\end{eqnarray}

Defining the new coordinates
\begin{eqnarray}
T=s,
\qquad
R=\frac{2r_0^{3/2}}{3\sqrt{2MG}},
\qquad
t=t(s),
\qquad
r=r(s),
\end{eqnarray}
we obtain the coordinate transformation
\begin{eqnarray}\label{TR}
T&=&t+2\sqrt{2MG\,r}
+2MG\ln\left|
\frac{\sqrt{r}-\sqrt{2MG}}{\sqrt{r}+\sqrt{2MG}}
\right| ,\nonumber\\
R&=&t+\frac{2}{\sqrt{2MG}}
\left(
\frac{r^{3/2}}{3}+2MG\sqrt{r}
\right)
+2MG\ln\left|
\frac{\sqrt{r}-\sqrt{2MG}}{\sqrt{r}+\sqrt{2MG}}
\right|.
\end{eqnarray}

This transformation coincides with formula (100.1) of Landau,
\begin{eqnarray}
T=t+\int\frac{r f(r)}{r-2MG}\,dr,
\qquad
R=t+\int\frac{r}{(r-2MG)f(r)}\,dr,
\end{eqnarray}
with
$f(r)=\sqrt{\frac{2MG}{r}}$.

From these expressions, we find
\begin{eqnarray}
R-T=\frac{2}{3\sqrt{2MG}}\,r^{3/2},
\end{eqnarray}
and the invariant line element in the new coordinates, which are the well-known Lemaître coordinates \cite{Lemaitre},  reads
\begin{eqnarray}
ds^2
=dT^2-
\left(
\frac{3}{4MG}(R-T)
\right)^{-2/3}dR^2
=dT^2-\frac{2MG}{r}\,dR^2 .
\end{eqnarray}

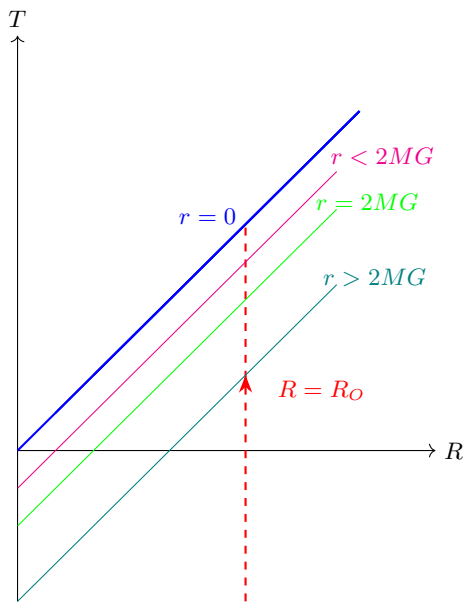
\begin{figure}[t]
\centering
\begin{tikzpicture}[scale=1.0]

\draw[->] (0,0) -- (5.5,0) node[right] {$R$};
\draw[->] (0,-2) -- (0,5.5) node[above] {$T$};

\usetikzlibrary{decorations.markings}

\tikzset{
  arrowcurve/.style={
    postaction={
     decorate,
      decoration={
        markings,
       mark=at position 0.6 with 
{\arrow[>=Stealth]{>}}       
      }
    }
  }
}

\draw[red, thick,dashed,arrowcurve] plot[smooth,domain=-2:3] (3, {\x});

\node[green] at (4.6,3.3) {$r=2MG$};

\node[teal] at (4.7,2.3) {$r>2MG$};

\node[magenta] at (4.8,3.9) {$r<2MG$};

\draw[blue, thick] (0,0) -- (4.5,4.5);
\draw[blue, thick] (0.5,0.5) -- (4.5,4.5);
\draw[blue, thick] (1,1) -- (4.5,4.5);

\node[blue] at (2.5,3.1) {$r=0$};

\draw[green] plot[smooth,domain=0:4.2] (\x, {(\x-1)});
\draw[teal] plot[smooth,domain=0:4.2] (\x, {(\x-2)});

\draw[magenta] plot[smooth,domain=0:4.2] (\x, {(\x-0.5)});

\node[red] at (4,0.8) {$R=R_O$};


\end{tikzpicture}
\caption{Representation of the Schwarzschild spacetime in Lemaître coordinates $(T,R)$. Freely falling worldline is represented by the dashed red line. The green straight-line corresponds to the horizon at $r=2MG$.}\label{figure1}
\end{figure}

In these coordinates, the worldlines of the freely falling particles
$(t(s),r(s))$ given in Eq.~(\ref{worldlines}) become straight lines
$(T,R)$ (see Figure \ref{figure1}).

Moreover, the invariant is singular only at $R-T=0$, which corresponds
to $r=0$. The line defined by
\begin{eqnarray}
\frac{3}{4MG}(R-T)=1
\end{eqnarray}
with finite $T$ and $R$ corresponds to $r=2MG$ and $t=+\infty$.


\bigskip
\noindent
{\bf Final remark.}


Consider now a worldline with $E=0$ that escapes from the horizon, namely
\begin{eqnarray}
\dot{r}=+\sqrt{\frac{2MG}{r}},
\qquad
\left(1-\frac{2MG}{r}\right)\dot{t}=1 .
\end{eqnarray}

In this case we have:
\begin{eqnarray}
    r(s)=(r_0^{3/2}+\frac{3}{2}\sqrt{2MG}s)^{2/3},
\end{eqnarray}
and we have $r(-s_{2MG})=2MG$. On the other hand,
\begin{eqnarray}
    t-t_0=\int_0^s\frac{d\tau}{A(r(\tau))}=\int_{r_0}^{r(s)}
    \frac{r}{r-2MG}\sqrt{\frac{r}{2MG}}dr,
\end{eqnarray}
and thus, at $s=-s_{\rm H}$ we have
$t(-s_{\rm H})=-\infty$. Therefore, 
using (\ref{TR}) we find
$T(-s_{\rm H})=-\infty$.

This means that,  in coordinates $(T,R)$, and using the time $T$, these geodesics, stars at $T=-\infty$ at the horizon and ends at $r=+\infty$ at time $T=\infty$. This shows that these coordinates are not geodesically complete, because in proper time, the geodesic {cannot} be extendend backwards in time to $r=0$.

\

What we will find in next section is a geodesically complete coordinate system

\subsubsection{Geodesics with negative energy ($E<0$)}

We denote by $R$ the maximal value of the radial coordinate $r$ along the
trajectory. Then, from the equation
\begin{eqnarray}
\dot{r}^2=\frac{2MG}{r}+2E,
\end{eqnarray}
we obtain
\begin{eqnarray}
E=-\frac{MG}{R}.
\end{eqnarray}
On the other hand, using once again the notation
$A(r)=1-\frac{2MG}{r}$, we have
\begin{eqnarray}\label{t}
A^2(r)\dot{t}^2=1+2E=A(R).
\end{eqnarray}

From this equation we see that $-\frac{1}{2}<E<0$, or equivalently
$R>2MG$.

We further impose the more restrictive and physical condition:
\begin{eqnarray}
A(r)\dot{t}=\sqrt{1+2E}.
\end{eqnarray}

With this choice, outside the horizon $r=2MG$ one has $\dot{t}>0$, i.e.
the Schwarzschild coordinate time $t$ (the time of an static observer) evolves in the same direction as the
proper time. On the contrary, inside the horizon one has $\dot{t}<0$,
and therefore in this region $t$ and $s$ evolve in opposite directions.

For $\dot{r}$ both signs must be considered,
\begin{eqnarray}
\dot{r}=\pm\sqrt{\frac{2MG}{r}+2E}.
\end{eqnarray}
For the positive sign the particle moves away from $r=0$, while for the
negative sign the particle falls towards $r=0$.

We choose initial conditions at proper time $s=0$ given by $(t,r)=(0,R)$.
The particle reaches the horizon with $\dot{t}>0$ and arrives there in a
finite proper time, since
\begin{eqnarray}
s=-\int_R^{r(s)}\frac{dr}{
\sqrt{\frac{2MG}{r}-\frac{2MG}{R}}
},
\end{eqnarray}
which is convergent for all $r(s)\leq R$.
However, the particle reaches the horizon at $t=+\infty$, because
\begin{eqnarray}
t=\sqrt{A(R)}\int_0^s\frac{d\tau}{A(r(\tau))}
=\sqrt{A(R)}\int_{r(s)}^R
\frac{dr}{|\dot{r}|A(r)}
=\sqrt{A(R)}\int_{r(s)}^R
\frac{dr}{
\sqrt{\frac{2MG}{r}-\frac{2MG}{R}}\,A(r)
},
\end{eqnarray}
and this integral diverges when $r(s)=2MG$.

The solution can be continued in proper time inside the horizon, where
$\dot{t}<0$, up to a finite value $s_{{\rm END}}$ at which the particle
reaches the point $(t_{{\rm END}}>0,r=0)$. The same occurs when one evolves
backwards in proper time. In this case, at proper time $-s_{{END}}$
the particle starts at the point $(-t_{{\rm END}},r=0)$, reaches the horizon
in finite proper time at $(t=-\infty,r=2MG)$, and finally arrives at
$(0,R)$ at proper time $s=0$.

Note that, 
when $-\frac{1}{2}\leq E\leq 0$, these geodesics foliate the Schwarzschild
spacetime. Note that the limiting case $E=-\frac{1}{2}\longleftrightarrow R=2MG$
will be better understood once we introduce Novikov coordinates.

\

We now introduce the notation
$f_u\equiv\partial f/\partial u$. Then
\begin{eqnarray}
dt=t_T\,dT+t_R\,dR,
\qquad
dr=r_T\,dT+r_R\,dR.
\end{eqnarray}

Therefore,
\begin{eqnarray}
ds^2=
(A(r)t_T^2-A^{-1}(r)r_T^2)dT^2
+2(A(r)t_Tt_R-A^{-1}(r)r_Tr_R)dTdR
+(A(r)t_R^2-A^{-1}(r)r_R^2)dR^2.
\end{eqnarray}

Using $\dot{t}=t_T$, $\dot{r}=r_T$, and the constraint
(\ref{constraint}), we obtain
\begin{eqnarray}
ds^2=dT^2
+2(A(r)t_Tt_R-A^{-1}(r)r_Tr_R)dTdR
+(A(r)t_R^2-A^{-1}(r)r_R^2)dR^2.
\end{eqnarray}

In a geodesic coordinate system one has
\begin{eqnarray}
g_{TR}=0,
\end{eqnarray}
which implies
\begin{eqnarray}
A^2(r)t_Tt_R=r_Tr_R.
\end{eqnarray}
From this relation one finds
\begin{eqnarray}
A^2(r)t_R^2
=r_R^2\frac{\frac{2MG}{r}+2E}{1+2E}.
\end{eqnarray}

Consequently,
\begin{eqnarray}
g_{RR}
=A^{-1}(r)\left(A^2(r)t_R^2-r_R^2\right)
=-\frac{r_R^2}{A(R)}.
\end{eqnarray}

Finally, the invariant line element in the variables $(T,R)$ is
\begin{eqnarray}
ds^2=dT^2-\frac{r_R^2}{A(R)}\,dR^2.
\end{eqnarray}

\

We consider solutions with initial conditions $(t(0),r(0))=(0,R)$.
A straightforward integration yields
\begin{eqnarray}
s=\pm\sqrt{\frac{R^3}{2MG}}
\left[
\arccos\left(\sqrt{\frac{r(s)}{R}}\right)
+\sqrt{\frac{r(s)}{R}}
\sqrt{1-\frac{r(s)}{R}}
\right].
\end{eqnarray}

An important fact should be emphasized: since the geodesics with $E<0$
satisfy condition (\ref{t}), they are only defined for $R\geq 2MG$.
Therefore, it is convenient to introduce the coordinate
\begin{eqnarray}
R=\frac{4M^2G^2+\bar{R}^2}{2MG},
\end{eqnarray}
which allows for an analytic extension.

In terms of these coordinates we obtain the relation
\begin{eqnarray}\label{T}
T^2=\frac{(\bar{R}^2+4M^2G^2)^3}{(2MG)^4}
\left[
\arccos\left(
\sqrt{\frac{2MGr}{\bar{R}^2+4M^2G^2}}
\right)
+\sqrt{\frac{2MGr}{\bar{R}^2+4M^2G^2}}
\sqrt{1-\frac{2MGr}{\bar{R}^2+4M^2G^2}}
\right]^2,
\end{eqnarray}
where $\bar{R}\geq0$, although the expression can be extended by symmetry
to all real values.

\

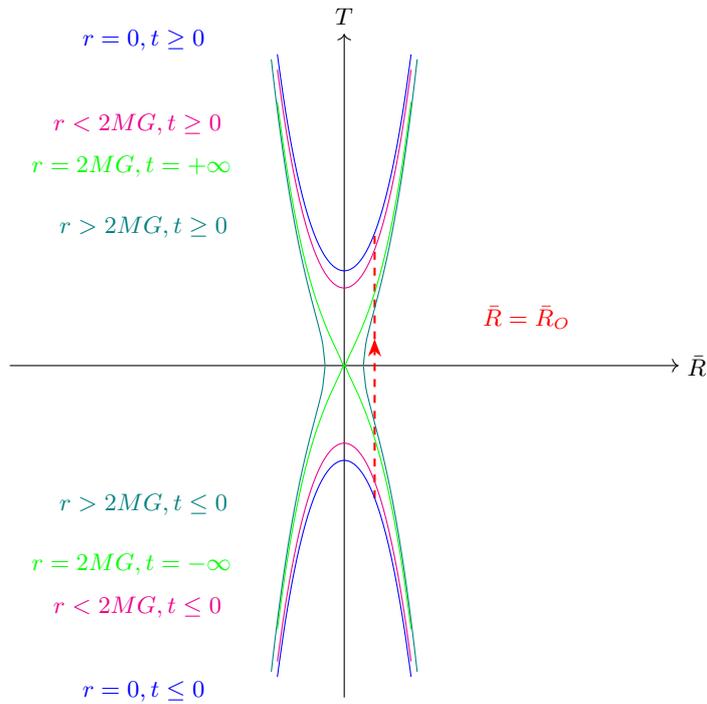
\begin{figure}[t]
\centering
\begin{tikzpicture}[scale=0.8]

\draw[->] (-5.5,0) -- (5.5,0) node[right] {$\bar{R}$};
\draw[->] (0,-5.5) -- (0,5.5) node[above] {$T$};

\usetikzlibrary{decorations.markings}

\tikzset{
  arrowcurve/.style={
    postaction={
     decorate,
      decoration={
        markings,
       mark=at position 0.6 with 
{\arrow[>=Stealth]{>}}       
      }
    }
  }
}

\draw[red, thick,dashed,arrowcurve] plot[smooth,domain=-2.2:2.2]
(0.5, {\x});

\node[green] at (-3.5,3.3) {$r=2MG, t=+\infty$};

\node[green] at (-3.5,-3.3) {$r=2MG, t=-\infty$};

\node[teal] at (-3.3,-2.3) {$r>2MG, t\leq 0$};

\node[teal] at (-3.3,2.3) {$r>2MG, t\geq  0$};

\node[magenta] at (-3.4,-4) {$r<2MG, t\leq 0$};

\node[magenta] at (-3.4,4) {$r<2MG, t\geq  0$};

\node[blue] at (-3.3,5.4) {$r=0, t\geq 0$};

\node[blue] at (-3.3,-5.4) {$r=0, t\leq 0$};

\node[red] at (3,0.8) {$\bar{R}=\bar{R}_O$};


\draw[blue]
plot[smooth,domain=-1.1:1.1] 
(\x, {(pi/2)*((\x)^2+1)^(3/2)});

\draw[blue]
plot[smooth,domain=-1.1:1.1] 
(\x, {-(pi/2)*((\x)^2+1)^(3/2)});

\draw[green]
plot[smooth,domain=-1.1:1.1] 
(\x, {((\x)^2+1)^(3/2)* (
acos(((\x)^2+1)^(-0.5))*pi/180+
((\x)^2)^(0.5)*((\x)^2+1)^(-1)});

\draw[green]
plot[smooth,domain=-1.1:1.1] 
(\x, {-((\x)^2+1)^(3/2)* (
acos(((\x)^2+1)^(-0.5))*pi/180+((\x)^2)^(0.5)*((\x)^2+1)^(-1)});

\draw[teal]
plot[smooth,domain=(0.1)^(0.5):1.2] 
(\x, { ((\x)^2+1)^(3/2)* (
acos((1.1)^(0.5)*((\x)^2+1)^(-0.5))*pi/180+(1.1)^(0.5)*((\x)^2-0.1)^(0.5)*((\x)^2+1)^(-1)});

\draw[teal]
plot[smooth,domain=(0.1)^(0.5):1.2] 
(\x, { -((\x)^2+1)^(3/2)* (
acos((1.1)^(0.5)*((\x)^2+1)^(-0.5))*pi/180+
(1.1)^(0.5)*((\x)^2-0.1)^(0.5)*((\x)^2+1)^(-1)});

\draw[teal]
plot[smooth,domain=-(0.1)^(0.5):-1.2] 
(\x, { ((\x)^2+1)^(3/2)* (
acos((1.1)^(0.5)*((\x)^2+1)^(-0.5))*pi/180+
(1.1)^(0.5)*((\x)^2-0.1)^(0.5)*((\x)^2+1)^(-1)});

\draw[teal]
plot[smooth,domain=-(0.1)^(0.5):-1.2] 
(\x, { -((\x)^2+1)^(3/2)* (
acos((1.1)^(0.5)*((\x)^2+1)^(-0.5))*pi/180+
(1.1)^(0.5)*((\x)^2-0.1)^(0.5)*((\x)^2+1)^(-1)});

\draw[magenta]
plot[smooth,domain=-(1.1):1.1] 
(\x, { ((\x)^2+1)^(3/2)* (
acos((0.5)^(0.5)*((\x)^2+1)^(-0.5))*pi/180+(0.5)^(0.5)*((\x)^2+0.5)^(0.5)*((\x)^2+1)^(-1)});

\draw[magenta]
plot[smooth,domain=-1.1:1.1] 
(\x, { -((\x)^2+1)^(3/2)* (
acos((0.5)^(0.5)*((\x)^2+1)^(-0.5))*pi/180+(0.5)^(0.5)*((\x)^2+0.5)^(0.5)*((\x)^2+1)^(-1)});

\end{tikzpicture}
\caption{Representation of the Schwarzschild spacetime in Novikov coordinates $(T,\bar{R})$. Freely falling worldline is represented by the dashed red line. }\label{figure2}
\end{figure}

These curves play a fundamental role (see Figure \ref{figure2}). We distinguish:

\begin{enumerate}
\item $r=0$. In this case
\begin{eqnarray}
T^2=\frac{\pi}{2}\,
\frac{(\bar{R}^2+4M^2G^2)^3}{(2MG)^4},
\end{eqnarray}
which has two branches. The positive branch corresponds in
Schwarzschild--Droste coordinates to $r=0$ and $t\geq0$, while the negative
branch corresponds to $r=0$ and $t\leq0$.

\item $r=2MG$. One obtains two curves corresponding to
$t=+\infty$ and $t=-\infty$, respectively.

\item $r=a>2MG$. Since
$\frac{2MGr}{\bar{R}^2+4M^2G^2}\leq1$, the minimal value of $\bar{R}^2$ is
$\bar{R}_{\min}^2=2MG(a-2MG)$. Hence the curve does not cross the axis
$\bar{R}=0$. Moreover, $T(\bar{R}_{\min})=0$.

\item $r=a<2MG$. In this case the curve crosses the axis $\bar{R}=0$, but
does not cross the axis $T=0$.
\end{enumerate}

In Novikov coordinates $(T,\bar{R})$ the invariant takes the form
\begin{eqnarray}
ds^2=dT^2-
\left(1+\frac{4M^2G^2}{\bar{R}^2}\right)
r_{\bar{R}}^2\,d\bar{R}^2,
\end{eqnarray}
where {$r_{\bar{R}}\equiv\frac{\partial r}{\partial \bar{R}}$} 
is obtained by differentiating {implicitly} Eq.~(\ref{T}) with respect
to $\bar{R}$, {keeping $T$ fixed, i.e. imposing $\frac{\partial T}{\partial \bar{R}}=0$.}

The geodesics $(T=s,\bar{R}\neq 0)$ correspond precisely to the geodesics
discussed above in Schwarzschild--Droste coordinates. 
What, then, is the meaning of the geodesic $(T=s,\bar{R}=0)$?

In Schwarzschild--Droste coordinates this curve must be understood as a
\emph{limiting trajectory}. It can be obtained as the limit of the
radial timelike geodesics when $E\to -1/2$. In this limit, the motion
degenerates as follows: the trajectory starts at $(t=0,r=0)$, follows
the line $(t=0,r)$ up to $(t=0,2MG)$, then evolves along the horizon
$r=2MG$ from $t=-\infty$ to $t=+\infty$, and finally returns to
$(t=0,r=0)$.

\

A final remark is in order:
Within this framework, the appearance of white-hole regions admits a clear and physically transparent interpretation.
In Schwarzschild--Droste coordinates, the time-reversed extension of the black-hole solution is often interpreted as a distinct physical object, namely a white hole, from which matter emerges.
However, this interpretation is largely coordinate-dependent.

In free-fall (Novikov-type) coordinates, the spacetime is naturally foliated by timelike geodesics describing the motion of freely falling observers.
In this description, the curves corresponding to the white-hole sector arise simply as the time-reversed continuation of ingoing geodesics across the singularity at $r=0$.
They do not represent an independent dynamical process, nor the explosive ejection of matter from a compact object, but rather the extension of the geodesic congruence beyond the classical singularity.

From this point of view, white holes should not be regarded as physically realizable objects formed by gravitational collapse.
Instead, they reflect the maximal analytic extension of the Schwarzschild geometry and the use of coordinates adapted to static observers at infinity.
For freely falling observers, the physically relevant description of gravitational collapse involves no horizon pathology and no observable white-hole behavior.

\begin{figure}[t]
\centering
\begin{tikzpicture}[scale=0.6]

\draw[->] (0,0) -- (5.5,0) node[right] {$r$};
\draw[->] (0,-5.5) -- (0,5.5) node[above] {$t$};
\draw[green] plot[smooth,domain=-6:6]
(2, {\x});

\node[green] at (3.5,0.8) {$r=2MG$};


\usetikzlibrary{decorations.markings}

\tikzset{
  arrowcurve/.style={
    postaction={
     decorate,
      decoration={
        markings,
       mark=at position 0.6 with 
{\arrow[>=Stealth]{<}}       
      }
    }
  }
}

\draw[red, thick,dashed,arrowcurve]
plot[smooth,domain= 2.012:3] 
 (\x, { -sqrt(2)*ln(\x-2)   });

\tikzset{
  arrowcurve/.style={
    postaction={
     decorate,
      decoration={
        markings,
       mark=at position 0.6 with 
       {\arrow[>=Stealth]{>}}          
      }
    }
  }
}

\draw[red, thick,dashed,arrowcurve]
plot[smooth,domain= 2.012:3] 
(\x, { sqrt(2)*ln(\x-2)   });

\tikzset{
  arrowcurve/.style={
    postaction={
     decorate,
      decoration={
        markings,
       mark=at position 0.6 with 
        {\arrow[>=Stealth]{>}}         
      }
    }
  }
}

\draw[red, thick,dashed, arrowcurve]
plot[smooth,domain= 0:1.95] 
(\x, { sqrt(2)*ln(2-\x) -2  });

\tikzset{
  arrowcurve/.style={
    postaction={
     decorate,
      decoration={
        markings,
       mark=at position 0.6 with 
        {\arrow[>=Stealth]{<}}         
      }
    }
  }
}

\draw[red, thick,dashed,arrowcurve]
plot[smooth,domain= 0:1.95] 
(\x, { -sqrt(2)*ln(2-\x) +2  });

\end{tikzpicture}
\caption{Schematic representation of the geodesic appearing in Figure \ref{figure2} in the Schwarzschild-Droste coordenates.}\label{figure3}
\end{figure}
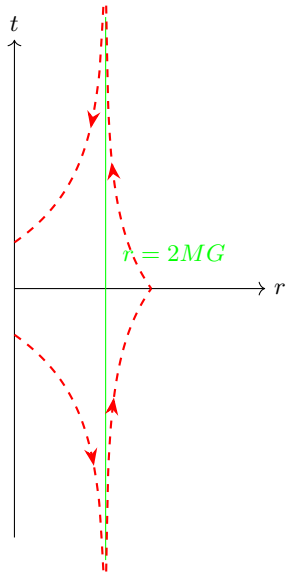


\section{Gravitational collapse}

Gravitational collapse was already considered shortly after the development of General Relativity, motivated by the search for relativistic stellar models and the understanding of the final stages of massive stars. Early investigations focused mainly on static or quasi-static configurations, such as relativistic stars in hydrostatic equilibrium \cite{Chadra}. However, it was only during the 1960s that the profound connection between gravitational collapse and the formation of spacetime singularities was clearly established. The seminal works of Penrose and Hawking demonstrated, under very general and physically reasonable assumptions, that singularities are an unavoidable feature of General Relativity once gravitational collapse leads to the formation of trapped surfaces \cite{Penrose, Hawking}. These results marked a turning point in our understanding of black holes, showing that singular behavior is not a consequence of idealized symmetry assumptions, but rather a generic prediction of the theory.

In this section, we consider a simple yet physically illuminating model of gravitational collapse, consisting of a spherically symmetric dust star, as originally analyzed by Oppenheimer and Snyder \cite{Oppenheimer}. Despite its simplicity, this model captures the essential features of gravitational collapse leading to black hole formation. Since the pressure vanishes identically, the conservation of the stress–energy tensor implies that each dust particle follows a timelike geodesic of the spacetime.
Effectively, one has \cite{Comer}:
\begin{equation}\label{stress-tensor}
\mathrm{div}(\mathfrak{T}) = 0 
\quad \Longleftrightarrow \quad 
\mathrm{div}(\rho {\bf u}) = 0 
\quad\text{and}\quad 
\frac{D{\bf u}}{ds} = 0,
\end{equation}
where ${\bf u}$ denotes the four-velocity field. The condition $\mathrm{div}(\rho {\bf u}) = 0$ represents mass conservation, while the equation $D{\bf u}/ds = 0$ explicitly states that fluid elements  move along geodesics, that is, in the absence of external non-gravitational forces. Both the divergence operator $\mathrm{div}$ and the covariant derivative $D/ds$ are defined with respect to the Levi-Civita connection.

As a consequence, the dynamics of the collapse can be naturally described in terms of freely falling observers, whose proper time provides a physically meaningful temporal parameter throughout the entire evolution, including the crossing of the Schwarzschild horizon.

The Oppenheimer–Snyder model thus offers a clear and conceptually transparent framework to analyze the formation of a black hole from regular initial data, and to contrast the behavior of different coordinate systems in describing the same physical process. In particular, it allows one to distinguish unambiguously between coordinate singularities and genuine physical effects, and to clarify the role played by time orientation and observer dependence in the interpretation of the collapsing geometry.

\

Assuming that the collapse is purely radial, the motion of the particles forming the boundary of the star is geodesic and reduces to the "Newtonian equation"
$$
\ddot r = -\frac{GM}{r^2},
$$
where $M$ is the total mass of the star, which, acording to Newton's theory,  can be treated as effectively concentrated at the center $r=0$. Moreover, by virtue of the {\it shell theorem} \cite{Newton}, the motion of a dust particle located inside the star satisfies
$
\ddot r = -\frac{GM(r)}{r^2},
$ 
where $M(r)$ denotes the mass enclosed within the sphere of radius $r$ whose boundary contains the particle.

Let $r_B$ denote the initial radius of the star in the Schwarzschild-Droste coordinates.
In Novikov coordinates, which are comoving with freely falling observers—for whom, in the spirit of D’Alembert’s principle, the inertial force exactly compensates the gravitational one—we choose the initial hypersurface $T=0$ so that the boundary of the star is located at the point $(T,\bar{R})=(0,\bar{R}_B)$.
The boundary then follows a timelike geodesic and reaches the Schwarzschild radius $r=2GM$ after a finite proper time $T_H(\bar{R}_B)$, which marks the formation of the horizon.



\subsection{External observers and the role of coordinates}

A natural question concerns how the gravitational collapse of the star is described by an observer located far away from it.

If one adopts Schwarzschild--Droste coordinates $(t,r)$, the worldline of the boundary of the star approaches the Schwarzschild radius $r=2MG$ asymptotically, reaching the horizon only at coordinate time
\begin{eqnarray}
    t\rightarrow +\infty.
\end{eqnarray}
From this perspective, the collapse appears to slow down indefinitely, giving rise to the well--known picture of a ``frozen star'', according to which the formation of the horizon is never completed in finite time.

However, it is essential to understand the physical meaning underlying this description. An observer who remains at rest at a fixed Schwarzschild radius $r=\mathrm{const}$ is not a freely falling observer. Even if located very far from the star, such an observer must be subject to a non--gravitational force that exactly compensates the gravitational attraction exerted by the mass $M$ of the star. Otherwise, the observer would inevitably fall towards the center.

Therefore, the use of Schwarzschild time implicitly assumes the presence of an external supporting force acting on the observer, whose role is to counterbalance gravity. Although this force may be arbitrarily small at sufficiently large distances, it must act continuously and over arbitrarily long times in order to keep the observer static.

From this point of view, describing gravitational collapse exclusively in Schwarzschild coordinates amounts to introducing, albeit implicitly, an external agent into the problem. In a simplified model of collapse that only includes the star and freely moving observers, such an external force has no natural place and its physical origin, duration, and dynamical backreaction should, in principle, be specified.

This clarifies that the ``frozen star'' picture is not a property of the collapse itself, but rather a consequence of describing the process from the viewpoint of observers who are artificially prevented from falling by external forces. In contrast, coordinate systems adapted to free--falling observers, such as Novikov coordinates, provide a description in which the formation of the horizon occurs at finite proper time, in agreement with the physical experience of freely falling observers.

\subsection{Novikov coordinates and the resolution of the paradox}

If one considers only the gravitational system formed by the star and an external observer, with no additional forces present, then both the star and the observer move along geodesics, independently of their initial separation.

In Novikov coordinates, this physical situation is described naturally. Suppose that at $T=0$ the boundary of the star is located at $(0,\bar{R}_B)$, while a distant observer is located at $(0,\bar{R}_O)$ with $\bar{R}_O \gg \bar{R}_B$. Since both follow freely falling trajectories, the observer sees the boundary of the star cross the horizon at the finite proper time $T_H(\bar{R}_B)$.

Thus, in Novikov coordinates, the apparent paradox of an infinite collapse time disappears. The ``frozen star'' picture is revealed to be a coordinate artifact associated with the use of Schwarzschild time, rather than a genuine physical effect.

\subsubsection{Gravitational collapse and the absence of white holes.}
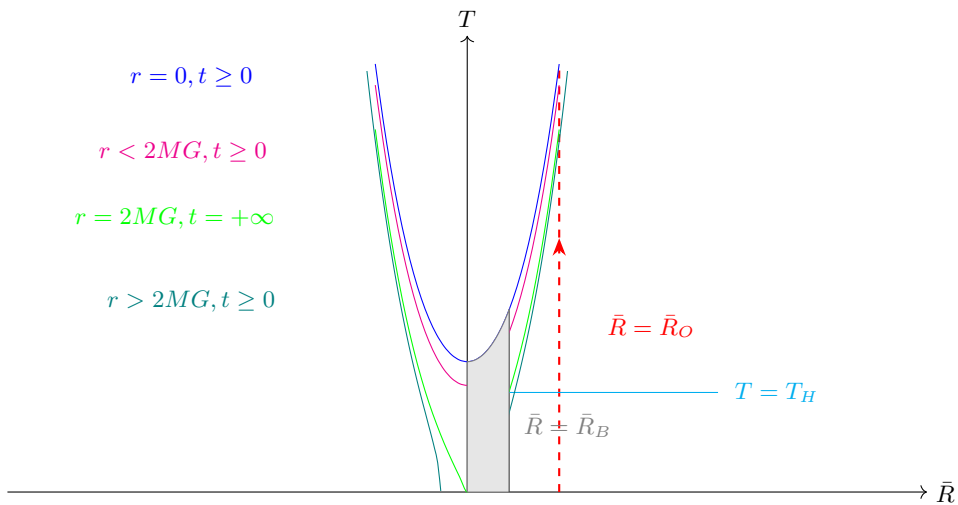
\begin{figure}[t]
\centering
\begin{tikzpicture}[scale=1.1]

\draw[->] (-5.5,0) -- (5.5,0) node[right] {$\bar{R}$};
\draw[->] (0,0) -- (0,5.5) node[above] {$T$};

\usetikzlibrary{decorations.markings}

\tikzset{
  arrowcurve/.style={
    postaction={
     decorate,
      decoration={
        markings,
       mark=at position 0.6 with 
{\arrow[>=Stealth]{>}}       
      }
    }
  }
}

\draw[red, thick,dashed,arrowcurve] plot[smooth,domain=0:5.1]
(1.1, {\x});

\draw[black] plot[smooth,domain=0:2.2]
(0.5, {\x});

\node[green] at (-3.5,3.3) {$r=2MG, t=+\infty$};

\node[teal] at (-3.3,2.3) {$r>2MG, t\geq  0$};

\node[magenta] at (-3.4,4.1) {$r<2MG, t\geq  0$};

\node[blue] at (-3.3,5) {$r=0, t\geq 0$};

\node[red] at (2.2,2) {$\bar{R}=\bar{R}_O$};

\node[gray] at (1.2,0.8) {$\bar{R}=\bar{R}_B$};

\node[cyan] at (3.7,1.2) {$T=T_H$};


\draw[blue]
plot[smooth,domain=-1.1:1.1] 
(\x, {(pi/2)*((\x)^2+1)^(3/2)});

\draw[green]
plot[smooth,domain=-1.1:1.1] 
(\x, {((\x)^2+1)^(3/2)* (
acos(((\x)^2+1)^(-0.5))*pi/180+
((\x)^2)^(0.5)*((\x)^2+1)^(-1)});

\draw[cyan]
plot[smooth,domain=0:3] 
(\x, {1.2});

\draw[teal]
plot[smooth,domain=(0.1)^(0.5):1.2] 
(\x, { ((\x)^2+1)^(3/2)* (
acos((1.1)^(0.5)*((\x)^2+1)^(-0.5))*pi/180+(1.1)^(0.5)*((\x)^2-0.1)^(0.5)*((\x)^2+1)^(-1)});

\draw[teal]
plot[smooth,domain=-(0.1)^(0.5):-1.2] 
(\x, { ((\x)^2+1)^(3/2)* (
acos((1.1)^(0.5)*((\x)^2+1)^(-0.5))*pi/180+
(1.1)^(0.5)*((\x)^2-0.1)^(0.5)*((\x)^2+1)^(-1)});

\draw[magenta]
plot[smooth,domain=-(1.1):1.1] 
(\x, { ((\x)^2+1)^(3/2)* (
acos((0.5)^(0.5)*((\x)^2+1)^(-0.5))*pi/180+(0.5)^(0.5)*((\x)^2+0.5)^(0.5)*((\x)^2+1)^(-1)});

\path[fill=gray!20, draw=gray]
(0,0)
-- plot[smooth,domain=0:0.5, samples=100]
   (\x, {0})
-- plot[smooth,domain=0.5:0, samples=100]
   (\x, {(pi/2)*((\x)^2+1)^(3/2)})
-- cycle;

\end{tikzpicture}
\caption{
Representation of gravitational collapse in Novikov coordinates $(T,\bar{R})$.
The star, whose evolution—restricted to those dust particles that at $T=0$ lie outside the Schwarzschild radius—is depicted by the gray region, exhausts its internal energy at $T=0$ and subsequently begins to collapse from an initial radius $R_B$, which corresponds to the Novikov coordinate
$\bar{R}_B=\sqrt{2MG,(R_B-2MG)}$.
$T_H$ is the time when the boundary crosses the Schwarzschild radius and the horizon is formed. $\bar{R}_O$ is the coordinate of an external observer following a geodesic with $-1/2<E<0$.}\label{figure5}
\end{figure}

When the Schwarzschild geometry is interpreted in the context of a realistic gravitational collapse, the notion of a white hole loses any physical meaning.
In an actual collapsing star, one has no access to, nor control over, the spacetime configuration prior to the formation of the star itself.
Consequently, the only physically meaningful description is the final stage of the evolution, when all internal sources of pressure and energy have been exhausted and the collapse becomes unavoidable
(see Figure \ref{figure5}).

In this regime, the motion of matter is well described by freely falling timelike geodesics, and the formation of the horizon is inevitable.
There is no physical process that could correspond to the time-reversed evolution associated with a white hole, namely the spontaneous emergence of matter from within the horizon.
Such a process would require finely tuned initial conditions imposed on a past singularity, which have no physical justification in the context of stellar evolution.

Therefore, in the description of gravitational collapse, white holes should be regarded as purely mathematical artifacts arising from the maximal analytic extension of the Schwarzschild solution.
They do not represent alternative physical outcomes of collapse, but rather reflect the use of global coordinate extensions that are not adapted to the causal structure of realistic astrophysical processes.

\section{Comparison with 
Kruskal–Szekeres coordinates}

The Kruskal--Szekeres coordinates \cite{Kruskal,Szekeres}
provide a global and purely geometric construction that removes the
coordinate singularity of the Schwarzschild--Droste metric at the
horizon $r=2MG$. These coordinates are designed so that the metric is
regular across the horizon and the maximal analytic extension of the
Schwarzschild spacetime becomes manifest.

The Kruskal--Szekeres coordinates, denoted by $(T,X)$, are related to the
Schwarzschild--Droste coordinates $(t,r)$ by

\begin{eqnarray}
    X^2-T^2=\left(\frac{r}{2MG}-1\right)e^{\frac{r}{2MG}},
    \qquad
    t=2MG\ln\left|\frac{X+T}{X-T}\right|.
\end{eqnarray}
In these coordinates the Schwarzschild line element becomes
\begin{eqnarray}
    ds^2=\frac{32M^3G^3}{r}e^{-\frac{r}{2MG}}\left(dT^2-dX^2\right).
\end{eqnarray}

\subsection*{Curves of constant $r$}

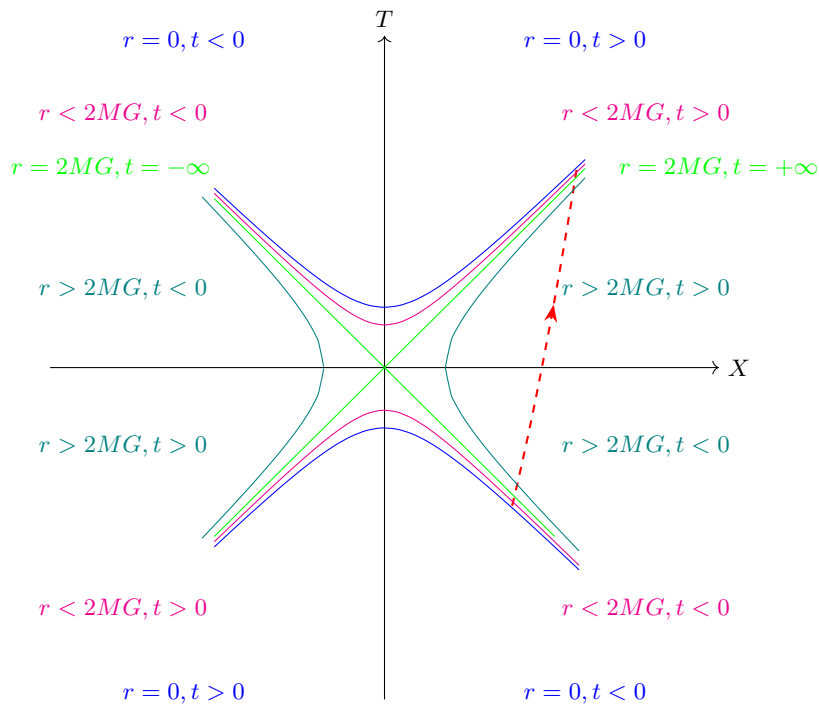
\begin{figure}[t]
\centering
\begin{tikzpicture}[scale=0.8]

\draw[->] (-5.5,0) -- (5.5,0) node[right] {$X$};
\draw[->] (0,-5.5) -- (0,5.5) node[above] {$T$};

\node[green] at (5.5,3.3) {$r=2MG, t=+\infty$};

\node[green] at (-4.5,3.3) {$r=2MG, t=-\infty$};

\node[teal] at (-4.3,-1.3) {$r>2MG, t> 0$};

\node[teal] at (4.3,1.3) {$r>2MG, t> 0$};

\node[teal] at (4.3,-1.3) {$r>2MG, t< 0$};

\node[teal] at (-4.3,1.3) {$r>2MG, t<  0$};

\node[magenta] at (-4.3,-4) {$r<2MG, t> 0$};

\node[magenta] at (-4.3,4.2) {$r<2MG, t< 0$};

\node[magenta] at (4.3,4.2) {$r<2MG, t> 0$};

\node[magenta] at (4.3,-4) {$r<2MG, t< 0$};

\node[blue] at (-3.3,5.4) {$r=0, t< 0$};

\node[blue] at (-3.3,-5.4) {$r=0, t> 0$};

\node[blue] at (3.3,-5.4) {$r=0, t< 0$};

\node[blue] at (3.3,5.4) {$r=0, t> 0$};


\usetikzlibrary{decorations.markings}

\tikzset{
  arrowcurve/.style={
    postaction={
     decorate,
      decoration={
        markings,
       mark=at position 0.6 with 
{\arrow[>=Stealth]{>}}       
      }
    }
  }
}

\draw[red, thick,dashed,arrowcurve] plot[smooth,domain= 2.1:3.16] (\x, { (\x)^2-6.7});

\draw[blue]
plot[smooth,domain=-2.8:3.3] 
(\x, {sqrt((\x)^2+1)});

\draw[blue]
plot[smooth,domain=-2.8:3.2] 
(\x, {-sqrt((\x)^2+1)});

\draw[magenta]
plot[smooth,domain=-2.8:3.3] 
(\x, {sqrt((\x)^2+0.5)});

\draw[magenta]
plot[smooth,domain=-2.8:3.2] 
(\x, {-sqrt((\x)^2+0.5)});

\draw[green]
plot[smooth,domain=-2.8:3.3] 
(\x, {((\x)});

\draw[green]
plot[smooth,domain=-2.8:2.8] 
(\x, {-(\x)});

\draw[teal]
plot[smooth,domain=1:3.2] 
(\x, {-sqrt((\x)^2-1)});

\draw[teal]
plot[smooth,domain=1:3.3] 
(\x, {sqrt((\x)^2-1)});

\draw[teal]
plot[smooth,domain=-3:-1] 
(\x, {-sqrt((\x)^2-1)});

\draw[teal]
plot[smooth,domain=-3:-1] 
(\x, {sqrt((\x)^2-1)});

\end{tikzpicture}
\caption{Schematic representation of a physical  geodesic with $-1/2<E<0$ in Kruskal spacetime. }\label{figure4}
\end{figure}

The curves $r=\text{const}$ are hyperbolae in the $(T,X)$ plane (see Figure \ref{figure4}).

\begin{enumerate}
\item \textbf{$r=0$.}  
The singularity corresponds to
\begin{eqnarray}
    T^2-X^2=1
    \qquad \mbox{and}\qquad 2MG\ln\left|\frac{X+T}{X-T}
    \right|=t.    
\end{eqnarray}
It consists of two branches symmetric under $(T,X)\to(-T,-X)$.  
The upper branch corresponds to $t>0$ for $X>0$ and $t<0$ for $X<0$, while the lower branch exhibits the opposite behavior.

\item \textbf{$r=2MG$.}  
The horizon is represented by the null lines
\begin{eqnarray}
    T=\pm X
    \qquad \mbox{and}\qquad 2MG\ln\left|\frac{X+T}{X-T}
    \right|=t    
    ,
\end{eqnarray}
corresponding respectively to $t=+\infty$ and $t=-\infty$.  
Again, the configuration is invariant under $(T,X)\to(-T,-X)$.

\item \textbf{$r>2MG$.}  
The hyperbolae lie in the regions $X>0$ and $X<0$.  
For $X>0$, one finds $t$ having the same sign as $T$, while for $X<0$ the Schwarzschild time $t$ has the opposite sign.

\item \textbf{$0<r<2MG$.}  
The hyperbolae lie in the regions $T>0$ and $T<0$, and a similar sign inversion occurs between $t$ and $X$.
\end{enumerate}

\subsection*{Physical interpretation}

Both Kruskal and Novikov coordinates provide an extension of the Schwarzschild spacetime beyond the horizon.  
However, their physical interpretation is markedly different.

In Kruskal coordinates the spacetime is symmetric under
\begin{eqnarray}
    (T,X)\longrightarrow(-T,-X),
\end{eqnarray}
and this symmetry allows two families of radial timelike geodesics:
\begin{eqnarray}
    A(r)\dot t=\pm\sqrt{1+2E},
    \qquad
    \dot r=\pm\sqrt{\frac{2MG}{r}+2E}.
\end{eqnarray}

In the region $X>0$, the geodesics satisfy $\dot t>0$ outside the horizon,
so the Schwarzschild time evolves in the same direction as the proper
time along the trajectory.  

In contrast, the region $X<0$ is filled by geodesics for which
$\dot t<0$ outside the horizon. In this case, the Schwarzschild time
coordinate evolves in the opposite direction to the proper time, so
these trajectories cannot be interpreted as physical motions of
particles as seen by static Schwarzschild observers. From the
Schwarzschild point of view, they would correspond to particles moving
backwards in coordinate time.  

Nevertheless, in Kruskal--Szekeres coordinates the time coordinate $T$
still increases monotonically along these geodesics, reflecting the fact
that the Kruskal construction remains a mathematically consistent and
regular extension of the spacetime.

In Novikov coordinates $(T,\bar{R})$ the situation is markedly different.
The spacetime is also doubled, now under the symmetry
$(T,\bar{R}) \rightarrow (T,-\bar{R})$, but \emph{both} regions are filled
by geodesics satisfying
\begin{eqnarray}
    A(r)\dot t=\sqrt{1+2E}.
\end{eqnarray}
Outside the horizon, all these geodesics satisfy $\dot t>0$, so the
Schwarzschild time evolves in the same direction as the proper time.
The Novikov time $T$ coincides with the proper time of free--falling
observers, and no time--reversed sector appears in the construction.

\

In conclusion, the region $X>0$ of the Kruskal spacetime and the region
$\bar{R}>0$ of the Novikov spacetime describe the same physical
Schwarzschild geometry. However, while the Novikov extension preserves
the physical time orientation everywhere, the Kruskal extension includes
an additional sector ($X<0$) corresponding to geodesics for which the
Schwarzschild time runs opposite to the proper time.  

This difference reflects the fundamentally geometric nature of the
Kruskal construction versus the dynamical character of the Novikov
coordinates, which are explicitly tied to the motion of freely falling
observers.

In summary, 
this analysis shows that maximal analytic extension does not necessarily
coincide with physical extension: while the former is dictated by purely
geometric criteria, the latter must respect the dynamical time
orientation selected by freely falling observers.



\section{Conclusions}

In this essay we have analyzed the Schwarzschild spacetime from the
perspective of coordinate systems adapted to different classes of
observers, with particular emphasis on the Novikov coordinates constructed
from a geodesic congruence of radial worldlines corresponding to freely
falling massive particles. By deriving these coordinates explicitly from
radial timelike geodesics, we have reviewed the well-known fact that the
Schwarzschild horizon is a regular hypersurface crossed in finite proper
time, and that its apparent singular behavior in Schwarzschild--Droste
coordinates is purely a coordinate artifact associated with static
observers.

The comparison between Schwarzschild--Droste, Kruskal--Szekeres, and
Novikov coordinates highlights the crucial role played by the choice of
time variable. While the Kruskal extension provides a mathematically
complete description of the eternal Schwarzschild solution, its physical
interpretation is not always transparent, since the Kruskal time is not
directly tied to the proper time—the true invariant in General Relativity—
of any physically distinguished class of observers. In contrast, the
Novikov time has a clear operational meaning, as it coincides with the
proper time along free--fall worldlines, and therefore offers a direct and
physically meaningful interpretation of the spacetime geometry {in situations where the motion is well approximated by radial geodesic
free fall}.

We have clarified the origin and meaning of trajectories that are often
interpreted as emerging from a white-hole region. Such trajectories arise
naturally in the maximally extended Schwarzschild spacetime as a result of
extending geodesics backwards in proper time, but they do not correspond to
physically realizable processes in scenarios involving gravitational
collapse {of spherically symmetric dust matter}. Once a
physically motivated time orientation is imposed—namely, that freely
falling observers evolve forward in proper time and that the Schwarzschild
time increases outside the horizon—the interpretation becomes unambiguous.
In this framework, no genuine white-hole behavior is required in collapsing
{dust} scenarios, as the past horizon is absent and only appears
in non-physical eternal constructions.

Furthermore, our analysis emphasizes that static observers in the
Schwarzschild geometry are necessarily non-inertial, since they must be
subject to a non-gravitational force that counterbalances the gravitational
attraction of the central mass. The Schwarzschild time coordinate is
adapted to this family of observers and coincides with their
proper time at fixed radius. The divergence of the Schwarzschild time at
the horizon therefore does not admit a dynamical explanation in terms of
the motion of freely falling particles, but rather reflects the breakdown
of the static observer congruence itself.
In contrast with the finite proper time measured by freely falling
observers, the apparent paradox is thus resolved by recognizing the
physical distinction between static and free--falling inertial observers.

Overall, the results presented here support the view that the essential
physics of {eternal} black holes is most naturally understood in
terms of free--fall motion and proper time. Coordinate systems adapted to
freely falling observers, such as the Novikov coordinates, provide a
transparent and physically meaningful description of the Schwarzschild
spacetime, and offer a valuable framework for discussing black hole
horizons without invoking unphysical artifacts {within the idealized setting of spherical symmetry and geodesic dust
motion}.

{Finally, we stress that the usefulness of Novikov coordinates in the
description of gravitational collapse is intrinsically tied to the
idealized setting considered throughout this work, namely that of
spherically symmetric configurations of pressureless matter, whose
constituents follow radial timelike geodesics. In more general collapse
scenarios, involving pressure, rotation, or deviations from spherical
symmetry, matter does not evolve along geodesic free--fall trajectories,
and coordinate systems adapted to a geodesic congruence lose their direct
dynamical relevance. In this sense, the Novikov framework should not be
viewed as a universal tool for realistic collapse modeling, but rather as
a pedagogically valuable and conceptually transparent framework for
clarifying observer--dependent aspects of horizon crossing, proper time,
and the interpretation of apparent paradoxes associated with black hole
spacetimes.}

\begin{acknowledgments}

JdH is supported by the Spanish grant PID2021-123903NB-I00
funded by MCIN/AEI/10.13039/501100011033 and by ``ERDF A way of making Europe''. 
\end{acknowledgments}

\end{document}